\documentclass[aps,prxquantum,twocolumn,superscriptaddress,longbibliography]{revtex4-2} 
\pdfoutput=1 % Math packages 
\usepackage{amsmath,amssymb,amsfonts} \usepackage{amsthm}\usepackage{physics} \usepackage{braket} \usepackage{bm} \usepackage{mathtools} 
\usepackage{amsmath}
\usepackage{graphicx} \usepackage{tikz} \usepackage{float} 
\usepackage{caption} \usepackage{subcaption} 
\usepackage{multirow} \usepackage{booktabs} 
\usepackage{algorithmicx} \usepackage{algpseudocode} 
\usepackage{xcolor} \usepackage{tcolorbox} 
\usepackage[english]{babel} % Hyperlinks 
\usepackage[colorlinks=true,linkcolor=blue,citecolor=blue,urlcolor=blue]{hyperref} 
\usepackage{enumitem} \usepackage{xcolor} \definecolor{criticalred}{HTML}{C0392B} \definecolor{highorange}{HTML}{E67E22} \definecolor{mediumpurple}{HTML}{8E44AD} \definecolor{lowgray}{HTML}{7F8C8D}     
\begin{document} 

\title{Trotterized hybrid qubit-boson simulation of the Jaynes-Cummings-Hubbard model} 
\twocolumngrid
\author{Nabaraj Khatri} 

\affiliation{Tri-Chandra Multiple Campus, Tribhuvan University, Ghantaghar, 44600 Kathmandu, Nepal} 
\affiliation{Holographic Himalaya, Lambagar, 44600 Tarakeshwar, Nepal}
\author{Dilip Ojha} 
\affiliation{Central Department of Physics, Tribhuvan University, Kirtipur, 44600 Kathmandu, Nepal}

\author{Rajib Tripathi} 
\affiliation{Central Department of Physics, Tribhuvan University, Kirtipur, 44600 Kathmandu, Nepal}

\author{Kismat K.C} 
\affiliation{Central Department of Physics, Tribhuvan University, Kirtipur, 44600 Kathmandu, Nepal}

\author{Ashish Khanal} 
\email{ak247626@ohio.edu} 
\affiliation{Department of Physics and Astronomy, Ohio University, Athens, OH 45701, USA} 
\affiliation{Holographic Himalaya, Lambagar, 44600 Tarakeshwar, Nepal}

\begin{abstract}
The Jaynes-Cummings-Hubbard (JCH) model describes the interaction of light and matter in coupled optical cavities and provides a framework for studying their quantum dynamics. In this work, we design a Trotterized quantum circuit to investigate real-time dynamics of JCH model using Bosonic Qiskit and a Matrix Product State (MPS) backend. We simulate single-polariton transport and benchmark site occupations against exact diagonalization. We then extend the simulations to the two-excitation sector under a hard-core photon constraint, examining excitation transport and symmetry properties. Finally, we study the dependence of bipartite entanglement entropy on the inter-cavity photon-hopping strength. The results illustrate how hybrid qubit--boson circuits, simulated with matrix product states, can be used to study light-matter lattice dynamics. This work provides a numerical benchmark for extending hybrid simulation approaches to more complex interacting and open light–matter systems.
\end{abstract}

\maketitle

% =====================================================================
\section{Introduction}\label{sec:intro}
% =====================================================================

Recent advances in quantum computing have led to the development of various computational frameworks and software tools for simulating quantum systems~\cite{feynman1982simulating,Lloyd1996,Chuang1998}. From circuit-based quantum simulators~\cite{javadiabhari2024quantum,bergholm2022pennylaneautomaticdifferentiationhybrid} to classical emulation and continuous-variable platforms~\cite{Lambert_2026,Killoran2019strawberryfields}, these tools provide ways to investigate quantum dynamics and explore the capabilities of emerging quantum hardware. However, efficiently simulating systems that combine discrete quantum degrees of freedom with bosonic modes remains a challenge due to the infinite-dimensional bosonic Hilbert space and boson-to-qubit encoding overhead~\cite{Macridin2018DigitalQC,Sawaya2020}. Such hybrid systems require computational frameworks capable of handling both qubit and bosonic degrees of freedom within a unified simulation environment~\cite{Stavenger2022,Liu_2026,s5bn-z4jq}.

A well known hybrid light--matter system is the Jaynes--Cummings--Hubbard (JCH) model, which couples two-level systems to continuous-variable cavity modes across a lattice to study polaritonic wave transport, quantum phase transitions, and spatial entanglement~\cite{Jaynes1963,Greentree_2006,Hartmann_2006,Noh2017}. Tracking its real-time dynamics is challenging due to infinite bosonic spaces and exponential state-space growth~\cite{CampagneIbarcq2020,Luchnikov2019}. Also, mapping continuous bosonic modes onto standard qubit gates introduces heavy circuit overhead~\cite{Sawaya2020}. Hence, we use Bosonic Qiskit~\cite{Stavenger2022}, a recently proposed Python package that provides built-in continuous-variable gates acting directly on truncated Fock-space qumodes alongside standard qubit operations~\cite{Macridin2018ElectronPhononSO,javadiabhari2024quantum}. We pair it with a Matrix Product State (MPS) tensor-network backend to keep memory usage low~\cite{Schollwck2011}. While Trotterized product-formula evolution has been explored for related spin--boson models~\cite{Trotter1959,Miessen2021QuantumAF,Chiesa2019}, Bosonic Qiskit has not yet been benchmarked on the many-body JCH chain against an exact diagonalization reference~\cite{Hauke2012,Weisse2008,Hoffmann2019}.

In this work, we use Bosonic Qiskit to simulate how a single polariton moves through a $1$D chain of $8$ connected cavities, each containing one cavity mode and one two-level atom coupled via local Jaynes--Cummings interactions and nearest-neighbor photon hopping~\cite{Jaynes1963}. Initializing the polariton at the middle cavity (site 4), we evolve the system via Trotterized local coupling and beam-splitter gates with time step $\Delta t = 0.05$ on an MPS backend. We benchmark the single-excitation dynamics against exact diagonalization, alongside an extension to two-excitation dynamics. We show that the polariton wavepacket spreads and reflects off the boundaries with finite-size revivals and the Trotter error remains small and bounded over several thousands steps. We also computed the bipartite entanglement entropy using same simulation tool and found that growth of entropy is governed by hopping strength ($J$) with fixed coupling strength($g$).

The rest of the paper is organized as follows. In Sec.~\ref{sec:cv-bosonic} we introduce Qiskit, continuous-variable quantum information and Bosonic Qiskit. In Sec.~\ref{sec:jch} we review the Jaynes--Cummings model and introduce the Jaynes--Cummings--Hubbard model. Section~\ref{sec:simulation} describes the simulation setup, the MPS backend, and the benchmark against exact diagonalization. Section~\ref{sec:entanglement} discusses the entanglement entropy dynamics, and Sec.~\ref{sec:conclusion} concludes with an outlook.

% =====================================================================
\section{Continuous-Variable Computing and Bosonic Qiskit}
\label{sec:cv-bosonic}
% =====================================================================

\subsection{Qubits and qumodes}
\label{sec:qubits-qumodes}

The fundamental unit of quantum information is the qubit, which, unlike a classical bit, can exist in a superposition of $|0\rangle$ and $|1\rangle$ \cite{divincenzo1995two}. IBM's Qiskit is an open-source software development kit (SDK) built to construct, simulate, and run quantum circuits on such qubits \cite{javadiabhari2024quantum, aleksandrowicz2019qiskit}. These gate-based frameworks are primarily built for individual qubit manipulation rather than complex coupled light--matter systems.

Many physical phenomena, such as electromagnetic field propagation, molecular vibration, and mechanical motion, are fundamentally continuous. In conventional discrete-variable (DV) platforms, these continuous observables must be mapped onto two-level qubit registers, which introduces substantial discretization overhead. Continuous-variable (CV) quantum computing avoids this mismatch by encoding information directly into the infinite-dimensional Hilbert space of a quantum harmonic oscillator (qumode), spanned by the Fock basis $\{|0\rangle, |1\rangle, |2\rangle, \dots \}$. A qumode's state is described equivalently by a pair of continuous, conjugate quadrature operators, $\hat{x}$ and $\hat{p}$, satisfying $[\hat{x}, \hat{p}] = i$ \cite{Braunstein2005}.

\subsection{Bosonic Qiskit}
\label{sec:bosonic-qiskit}

Bosonic Qiskit is a free, open-source Python tool made by the Co-design Center for Quantum Advantage (C2QA) as an add-on to IBM's Qiskit (we refer to read this for more details \cite{Stavenger2022}). It complements conventional quantum circuit models by introducing primitives for CV modes and hybrid qubit--boson architectures \cite{Bravyi2022}, and gives users a high-level framework to simulate, control, and analyze them . It directly incorporates bosonic creation/annihilation operators, displacement/squeezing dynamics, and multi-mode interactions alongside standard qubit registers, significantly reducing the overhead of discretizing infinite-dimensional Hilbert spaces.

To perform classical simulations, Bosonic Qiskit truncates each qumode to a finite Fock space $\{|0\rangle, |1\rangle, \dots, |N-1\rangle\}$, where $N$ is the chosen cutoff dimension. Operations on qumodes are governed by creation ($\hat{a}^\dagger$) and annihilation ($\hat{a}$) operators satisfying $[\hat{a}, \hat{a}^\dagger] = 1$, and are executed through unitaries such as the displacement operator $\hat{D}(\alpha) = \exp(\alpha \hat{a}^\dagger - \alpha^* \hat{a})$, conditional gates such as controlled displacements, and beam-splitter interactions $\hat{B}(\theta)$ between qumodes.

The total Hilbert space of a hybrid circuit with $m$ qubits and $n$ qumodes is the tensor product,
\begin{equation}
\mathcal{H}_{\text{hybrid}} = \mathcal{H}_{\text{qubit}}^{\otimes m} \otimes \mathcal{H}_{\text{qumode}}^{\otimes n} = \mathbb{C}^{2^m} \otimes \text{span}\left(\big\{|k\rangle\big\}_{k=0}^{N-1}\right)^{\otimes n}.
\end{equation}
Time evolution is implemented through matrix-exponential gates or explicit Hamiltonian simulation terms, such as the conditional rotation $\hat{U}_{\text{CR}}(\theta) = \exp\left(-i \theta \hat{\sigma}_z \hat{a}^\dagger \hat{a}\right)$, where $\hat{a}^\dagger \hat{a} = \hat{n}$ is the photon number operator and $\hat{\sigma}_z$ acts on the control qubit. Table~\ref{tab:gates} lists some of the gates available in Bosonic Qiskit.

\begin{figure}[!t]
    \centering
    \includegraphics[width=0.50\textwidth]{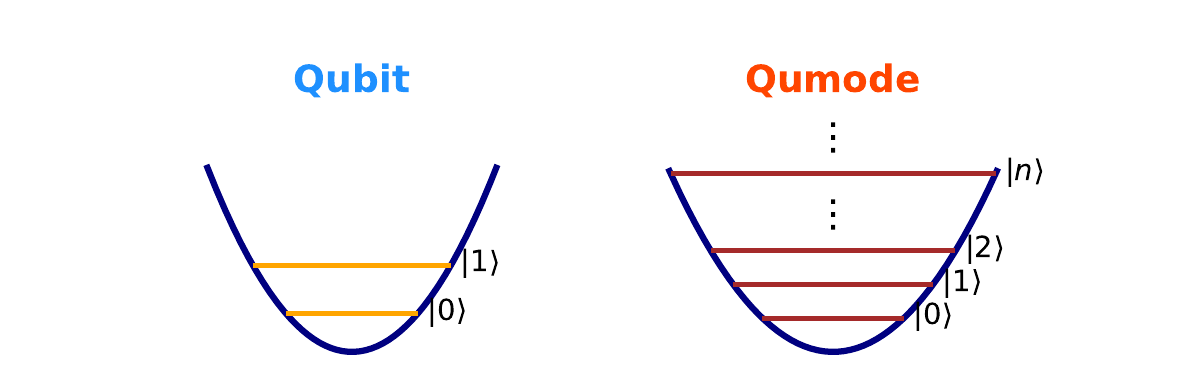}
    \caption{Comparison of potential energy wells and energy levels between a discrete qubit system ($\ket{0}, \ket{1}$) and a continuous-variable qumode system ($\ket{n}$), highlighting the transition between finite and infinite-dimensional Hilbert spaces in hybrid quantum architectures.}
    \label{fig:qubit_qumode}
\end{figure}

\begin{table}[t]
\begin{ruledtabular}
\begin{tabular}{lll}
Gate & Generator & Bosonic Qiskit call \\
\hline
Phase-space rotation & $e^{i\theta \hat n}$ & \texttt{cv\_r()} \\
Beamsplitter & $e^{\theta \hat a^\dagger \hat b - \theta^* \hat b^\dagger \hat a}$ & \texttt{cv\_bs()} \\
Single-mode squeezing & $e^{\frac12(\theta^*\hat a^2 - \theta \hat a^{\dagger 2})}$ & \texttt{cv\_sq()} \\
Controlled rotation & $e^{\sigma^z \otimes i\theta \hat n}$ & \texttt{cv\_c\_r()} \\
(Controlled) SNAP & $e^{\sigma^z\otimes i\theta_n \ket{n}\!\bra{n}}$ & \texttt{cv\_snap()} \\
\end{tabular}
\end{ruledtabular}
\caption{Selected continuous-variable gates in Bosonic Qiskit.}
\label{tab:gates}
\end{table}

To simulate the JCH Hamiltonian, one must account for both local spin--photon couplings and inter-site bosonic hopping. In later sections, we use a Trotterized gate scheme combining Jaynes--Cummings interaction unitaries with beam-splitter hopping networks.

% =====================================================================
\section{Jaynes--Cummings--Hubbard Model}
\label{sec:jch}
% =====================================================================

\subsection{Jaynes--Cummings model}
\label{sec:jcm}
The Jaynes--Cummings model (JCM) is a fully quantum mechanical model describing a single-mode field interacting with a two-level atomic system \cite{Jaynes1963,Shore1993}. It is like a mirrored cavity in which photons bounce back and forth between the mirrors. Although an atom has more than two energy levels, it is possible to confine transitions to only two specific quantum states of the atom, forming an effective two-level system.

To describe this two-level atom, we introduce the usual atomic raising and lowering operators $\sigma_+ = |e\rangle\langle g|$ and $\sigma_- = |g\rangle\langle e|$, along with the inversion operator $\sigma_z = |e\rangle\langle e| - |g\rangle\langle g|$, which together obey the standard Pauli algebra. The electric dipole interaction between the atom and the field then contains four terms, $\hbar g(\sigma_+ a + \sigma_+ a^\dagger + \sigma_- a + \sigma_- a^\dagger)$, where $g$ is the atom--field coupling strength. Of these, $\sigma_+ a^\dagger$ and $\sigma_- a$ do not conserve energy and are dropped under the rotating wave approximation (RWA), leaving only the energy-conserving terms $\sigma_+ a$ and $\sigma_- a^\dagger$. This gives the Jaynes--Cummings Hamiltonian,
\begin{equation}
H_{\text{JC}} = \frac{\hbar\omega_a}{2}\sigma_z + \hbar\omega_c a^\dagger a + \hbar g (\sigma_+ a + \sigma_- a^\dagger),
\end{equation}
where $\omega_a$ and $\omega_c$ are the atomic and cavity frequencies. It is often referred to as the ``hydrogen atom of quantum optics'' for its simplicity \cite{Shore1993}.

Working in the interaction picture and imposing resonance ($\omega_a = \omega_c$), the free part $H_0$ and the coupling $H'$ commute, so $H_I = H' = \hbar g(\sigma_+ a + \sigma_- a^\dagger)$. For a single excitation initially on the atom, $|\psi(0)\rangle = |e,n\rangle$, solving the resulting two-level Schr\"odinger equation gives the exact state,
\begin{equation}
|\psi(t)\rangle = \cos\!\big(g\sqrt{n+1}\,t\big)\,|e,n\rangle \;-\; i\sin\!\big(g\sqrt{n+1}\,t\big)\,|g,n+1\rangle,
\end{equation}
i.e. a coherent, periodic exchange of the excitation between the atom and the field mode. The atomic inversion $w(t) = \langle\hat\sigma_z\rangle = P_i(t) - P_f(t)$ then follows
\begin{equation}
w(t) = \cos(\Omega(n)\,t), \qquad \Omega(n) = 2g\sqrt{n+1},
\end{equation}
where $\Omega(n)$ is the Rabi frequency. In the vacuum ($n=0$) this reduces to $\Omega(0) = 2g$, the vacuum Rabi oscillation, in which the atom exchanges its excitation with the field even with no photons present initially.

\begin{figure*}[!t]
    \centering
    \includegraphics[width=\textwidth]{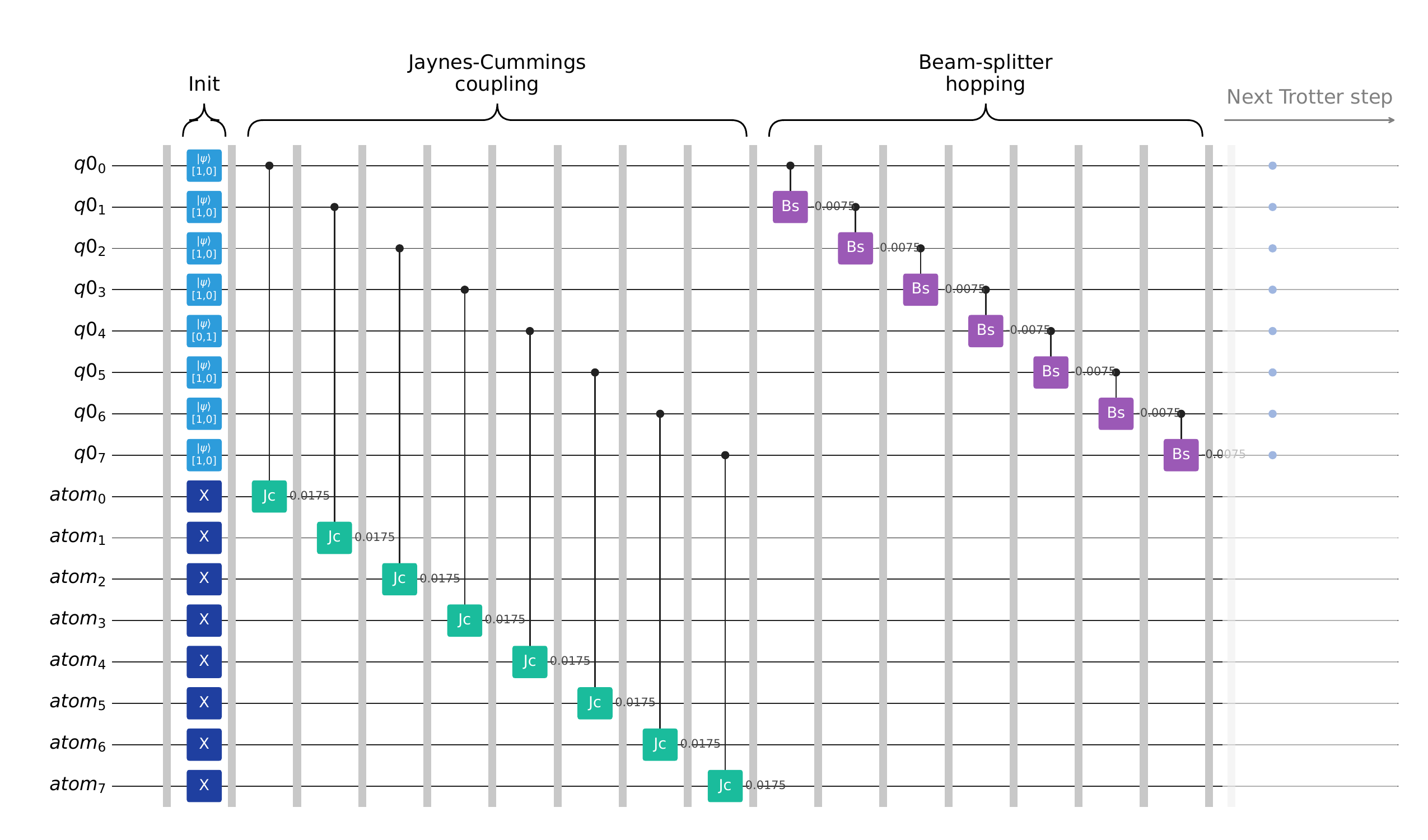}
    \caption{Jaynes–Cummings–Hubbard system implementation circuit setup with 8 sites for the first Trotter step of step size 0.05. The fourth site is initialized in the excited state, while the others are initialized in the ground state. The elongated vertical gates are Jaynes–Cummings gates which couples the photon atom interaction. The BS gate is the beam splitter that introduces the hopping and allow bosonic excitations to interfere and become distributed between modes. The jc coupling strength and hopping parameter are set to 0.35 and 0.15 respectively. The Pauli x gates set atoms in ground state.}
    \label{fig:jch_probe_circuit}
\end{figure*}
\subsection{Jaynes--Cummings--Hubbard Hamiltonian}
\label{sec:jch-lattice}

The JCH model combines Jaynes--Cummings cavity physics with inter-cavity photon hopping on a lattice \cite{Greentree_2006,Hartmann_2006}. Each site hosts a cavity mode coupled to a two-level atom, and photons can hop between neighboring sites. Setting $\hbar = 1$, its Hamiltonian reads,
\begin{equation}
H_{\text{JCH}} = -J \sum_{\langle i, j \rangle} (a_i^\dagger a_j + \text{h.c.}) + \sum_i h_i^{\text{JC}} - \mu \sum_i N_i,
\label{eq:JCH}
\end{equation}
\begin{equation}
h_i^{\text{JC}} = \omega_c a_i^\dagger a_i + \omega_a \sigma_i^+ \sigma_i^- + g(a_i^\dagger \sigma_i^- + a_i \sigma_i^+),
\end{equation}
\begin{equation}
N_i = a_i^\dagger a_i + \sigma_i^+ \sigma_i^-,
\end{equation}
where $J$ is the inter-cavity hopping rate, $g$ is the vacuum Rabi coupling, and $N_i$ is the total excitation number at site $i$ \cite{Greentree_2006}.

% ---- new block starts here ----
In our work we set $\mu=0$ and take the atoms and cavities on resonance, $\omega_c=\omega_a\equiv\omega$, on an open chain where $\langle i,j\rangle$ runs over nearest neighbors. The Hamiltonian conserves the total excitation number $\sum_i N_i$, so a state prepared with a single excitation stays in the single-excitation sector. There the term $\omega\sum_i N_i$ only contributes a global phase, and the dynamics are generated by,
\begin{equation}
H = -J \sum_i (a_i^\dagger a_{i+1} + \text{h.c.}) + g \sum_i (a_i^\dagger \sigma_i^- + a_i \sigma_i^+).
\label{eq:simulation}
\end{equation}
This is the Hamiltonian that we Trotterize and map onto qubit--qumode gates (JC and beam-splitter terms) in Sec.~\ref{sec:jch-setup}. Each site has the polariton states $|n,\pm\rangle = (|g,n\rangle \pm |e,n-1\rangle)/\sqrt{2}$ with energies $E_{n,\pm} = n\omega \pm g\sqrt{n}$. The initial state $|e,0\rangle$ is an equal superposition of $|1,+\rangle$ and $|1,-\rangle$, which is the origin of the vacuum Rabi oscillation at frequency $2g$.
% ---- new block ends here ----
\begin{figure*}[!t]
    \centering
    \includegraphics[width=1\textwidth]{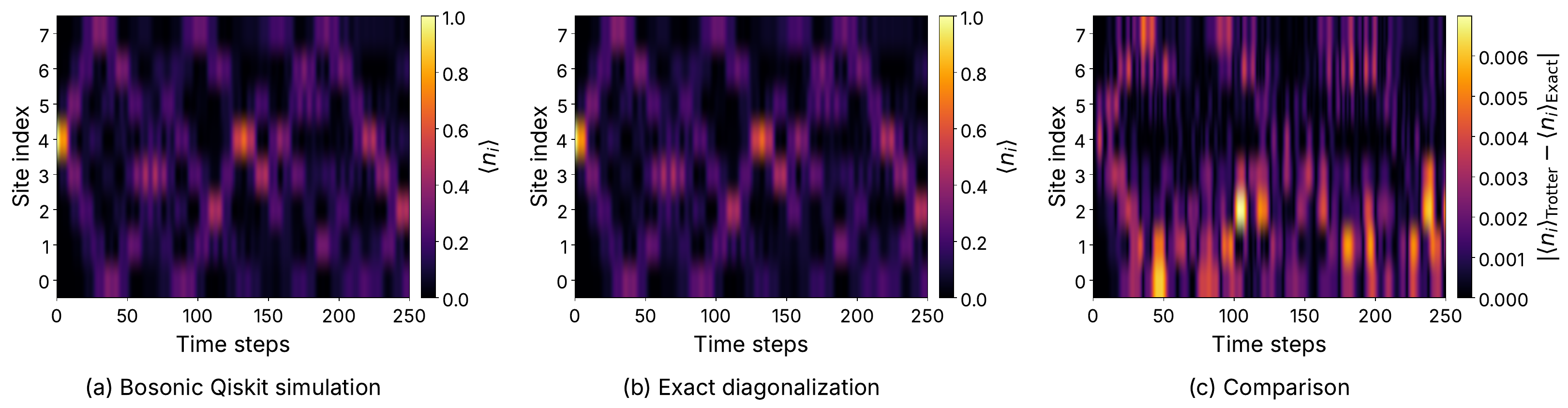}
    \caption{Time evolution of the site occupation $\langle n_i(t)\rangle$ for a single polariton on an $N=8$ JCH chain with open boundaries, $\omega_c=\omega_a$, $g=0.35$ and $J=0.15$, initialized at the central site ($i=4$). (a) Bosonic Qiskit Trotter circuit with Fock cutoff $d=2$ per qumode (matrix-product-state backend), run to $t=250$ with $\Delta t=0.05$ (5000 steps). (b) Exact diagonalization for the same setup. (c) Absolute difference between (a) and (b), with a maximum of $6.98\times10^{-3}$, consistent with ordinary Trotter error. The excitation spreads ballistically for $t\lesssim 20$, and reflections from the boundaries produce partial revivals near $t\approx130$ and $t\approx220$, the finite-size Poincar\'e-recurrence behavior.}
    \label{fig:Heatmap}
\end{figure*}

Photon blockade at each site gives the JCH model its Bose-Hubbard like character, the single-excitation and two-excitation polariton levels $E_{n,\pm}$ are non-degenerate, with an effective on-site repulsion,
\begin{equation}
U_{\text{eff}} = E_{2,-} - 2E_{1,-} = g(2-\sqrt{2}),
\end{equation}
which suppresses a second excitation from occupying the same cavity \cite{PhysRevA.76.031805,Greentree_2006}. This is what allows the JCH lattice to support a Mott-insulator-to-superfluid-like transition. However, we limit our work just for simulating the JCH cavity site occupancies and entanglement entropy using Bosonic qiskit.

% =====================================================================
\section{Trotterized Simulation of the JCH Chain}
\label{sec:simulation}
% =====================================================================

Having introduced Bosonic Qiskit's continuous-variable primitives in
Sec.~\ref{sec:cv-bosonic}, we now put them to work on the JCH Hamiltonian of Eq.~\eqref{eq:simulation}. We build
the Trotterized circuit, run it on a matrix-product-state backend, and
check the result against an independent exact solution for a single
excitation. Firstly, we establish that our implementation is correct by comparing with exact diagonalization. Second, we simulate two-excitation dynamics using a similar Trotter circuit. All simulations were performed in Python using Qiskit (v2.0.0), Qiskit Aer (v0.17.0), and Bosonic Qiskit (v15.1)~\cite{Stavenger2022}. The simulation code used in this work is available upon request.

\subsection{Circuit construction and initial state}
\label{sec:jch-setup}

We work with a one-dimensional JCH chain of $N=8$ sites with open
boundaries. Each site carries one qumode (the cavity) and one qubit
(the atom), so the total Hilbert space takes the hybrid form,
$\mathcal{H}_{\rm hybrid}
    = \mathcal{H}_{\rm qubit}^{\otimes m} \otimes \mathcal{H}_{\rm qumode}^{\otimes n}
    = \mathbb{C}^{2^m} \otimes \left(\mathrm{span}\{|k\rangle\}_{k=0}^{d-1}\right)^{\otimes n}$,
with $m=n=8$. Each qumode is truncated to a Fock cutoff $d=2$, so a cavity
can hold either zero or one photon.

It helps to picture what the model describes. Every site is a small
cavity with an atom inside it. When the two are on resonance, an
excitation keeps swapping between the cavity photon and the atom, so
at any moment it is partly light and partly matter. Such a mixed
excitation is called a polariton. The photon can also leak into the
neighboring cavity at rate $J$, while the atom cannot move. A polariton
therefore travels along the chain only through its photon part. This
is why the local excitation number that we track is the sum of the two
pieces,
\begin{equation}
    \hat N_i = a_i^\dagger a_i + \sigma_i^{+}\sigma_i^{-}.
    \label{eq:Ni}
\end{equation}
The on-site Jaynes--Cummings term only moves an excitation between the
photon and the atom of the same site, so it leaves $\hat N_i$ unchanged.
The fast photon--atom exchange is therefore invisible in $\langle
n_i\rangle$, and what we see in the occupancy is purely the motion of
polaritons from site to site. Both terms of the Hamiltonian also conserve
the total excitation number, which we use as a check later on.

In the single-excitation run we start the chain with one polariton at $i=4$. The qubit there is flipped to $|e\rangle$ with a Pauli-$X$ gate, every other atom is left in the ground state, and every qumode starts in vacuum. 

The Trotter step alternates two layers, matching the split of $H_{\rm JCH}$ into its on-site and hopping pieces. Firstly there is a layer of Jayne-Cummings gates, one per site, coupling each atom to its own
cavity with $g=0.35$. This exchanges an excitation
between photon and atom. Then comes a layer of beam-splitter gates
connecting neighboring qumodes, with hopping strength $J=0.15$, which
lets a photon hop from one cavity to the next. Since the atoms and
cavities are on resonance, the detuning is zero and no extra
single-mode phase gates are needed. Because these two layers do not
commute, splitting the evolution this way is only approximate, and the
error per step shrinks as $\Delta t$ is made smaller.
Figure~\ref{fig:jch_probe_circuit} shows this decomposition for the
first step, at $\Delta t = 0.05$.

We stop the circuit at a sequence of checkpoints out to $t=250$ (5000
Trotter steps), and at each one calculate the reduced density matrix
$\rho_i$ for every site and evaluate
\begin{equation}
    \langle n_i(t)\rangle = \mathrm{Tr}\!\left[\rho_i \hat N_i\right],
    \label{eq:occ-def}
\end{equation}
where $\hat N_i$ is given in Eq.~\eqref{eq:Ni}. Stacking
$\langle n_i(t)\rangle$ across all eight sites and every checkpoint
gives the space--time occupancy map, which we show as a heatmap in the
left panel of Fig.~\ref{fig:Heatmap}.\\

\begin{figure}[!t]
    \centering
    \includegraphics[width=1\linewidth]{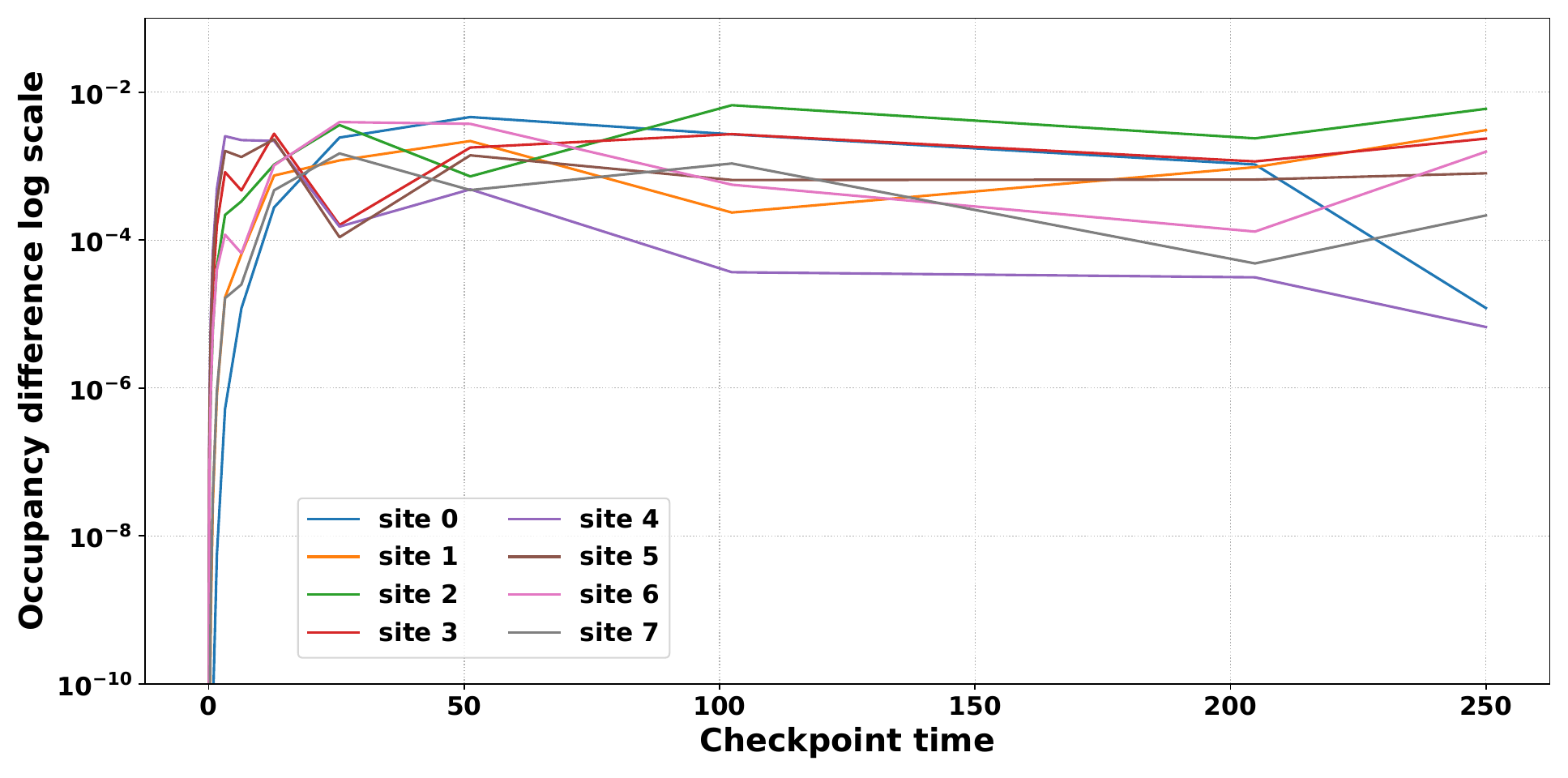}
    \caption{Site-wise error analysis of the Trotterized JCH dynamics against exact diagonalization for a single initial excitation (at the central site). For each site, the absolute occupation discrepancy $\lvert \langle n_i\rangle_{\rm Trotter}-\langle n_i\rangle_{\rm Exact}\rvert$ is plotted as a function of time at selected checkpoints. The vertical axis is logarithmic and the inset gives a magnified view. The residual deviations are consistent with finite-step Trotter error and confirm that the quantum-circuit evolution tracks the exact dynamics with controlled accuracy over the simulated window.}
    \label{fig:jch_occupancy_error}
\end{figure}
\subsection{Matrix Product State}
A dense statevector over the hybrid register scales as the product of
every local dimension, here $2^8 \times d^8$, so both memory and gate
cost grow exponentially in the number of sites even at a modest
cutoff. An MPS sidesteps this by writing the state as a chain of small
tensors linked by bond indices. The bond dimension $\chi$ sets how
much entanglement this chain can hold, and after every two-site gate
an SVD truncates the state back down, keeping only the $\chi$ largest
singular values \cite{PhysRevLett.91.147902, RevModPhys.77.259}.
We run the circuit with the matrix-product-state method of Qiskit Aer,
which is selected through a single backend option, so the circuit is
the same one that would run on a statevector backend.

That truncation is only cheap if the state does not need much
entanglement to describe. For a nearest-neighbor chain like JCH it
mostly does not. Since $H_{\rm JCH}$ only couples adjacent sites, the
entanglement generated during the dynamics tends to follow close to an
area-law scaling across the chain, rather than growing extensively
with every bipartition. This makes the MPS a suitable backend for the
JCH chain, and the same code can be used for longer chains.

\subsection{Benchmarking against exact diagonalization}
\label{sec:jch-benchmark}

To check the error, we compare the Trotterized MPS run against an exact solution for similar setup. Since our initial state carries exactly one excitation and the JCH Hamiltonian conserves excitation number, $H_{\rm JCH}$ restricted to this sector is an ordinary $2N \times 2N$ matrix in the basis of ``photon at site $i$'' and ``atom excited at site $i$'' states. It can be easily diagonalized which gives an exact reference.

In this sector the problem is a single particle hopping on a chain, which can sit on each site either as a photon or as an atom. The heatmaps in Fig.~\ref{fig:Heatmap} show the result. At $t=0$ the excitation sits on the central site and spreads outward, with the value of $J$ and slowed because only the photon part of the polariton can hop. Since $g>J$, the two polariton branches are separated by about $2g$, and their beating produces the fine streaks along the time axis. After reflecting off the open boundaries, the fronts return and interfere, giving partial revivals on the central site near $t\approx130$ and $t\approx220$.

The two calculations agree well and reproduce the same spreading, reflection, and revival pattern. The absolute difference peaks at $6.98\times10^{-3}$, about what one would expect from Trotter error at $\Delta t = 0.05$ rather than a circuit bug or an under-resolved bond dimension. Figure~\ref{fig:jch_occupancy_error} shows this site by site. The error grows during the initial ballistic phase, then saturates and fluctuates within a narrow band ($10^{-4}$ to $10^{-2}$) instead of accumulating, across all eight sites.\

\subsection{Two-excitation dynamics}
\label{sec:jch-multi}

 In this section we move beyond from single excitation and simulate two excitations. Each atom is a two-level system and each cavity is truncated to at most one photon. Hence, the site cannot absorb excitations without limit and the excitations effectively feel each other. Note that this excludes states like $|g,2\rangle$, so what we simulate is a truncated, hard-core version of the JCH model. The state space also grows quickly with the number of excitations, and we test the same circuit in this regime.

We take a chain of $N=7$ sites with cutoff $d=2$, which
gives $14$ qubits in total. The couplings are $g=0.35$ for the on-site
Jaynes--Cummings term and a stronger hopping $J=0.5$, so that photons now
move more easily than in the single-excitation run. Two photons are
placed on sites $2$ and $4$, symmetrically around the central site, while all atoms start in the ground state and the remaining
qumodes in vacuum. The Trotter step is the similar as before, with
$\Delta t=0.05$ and $3000$ steps, reaching $t=150$.

\begin{figure}[!t]
    \centering
    \includegraphics[width=1\linewidth]{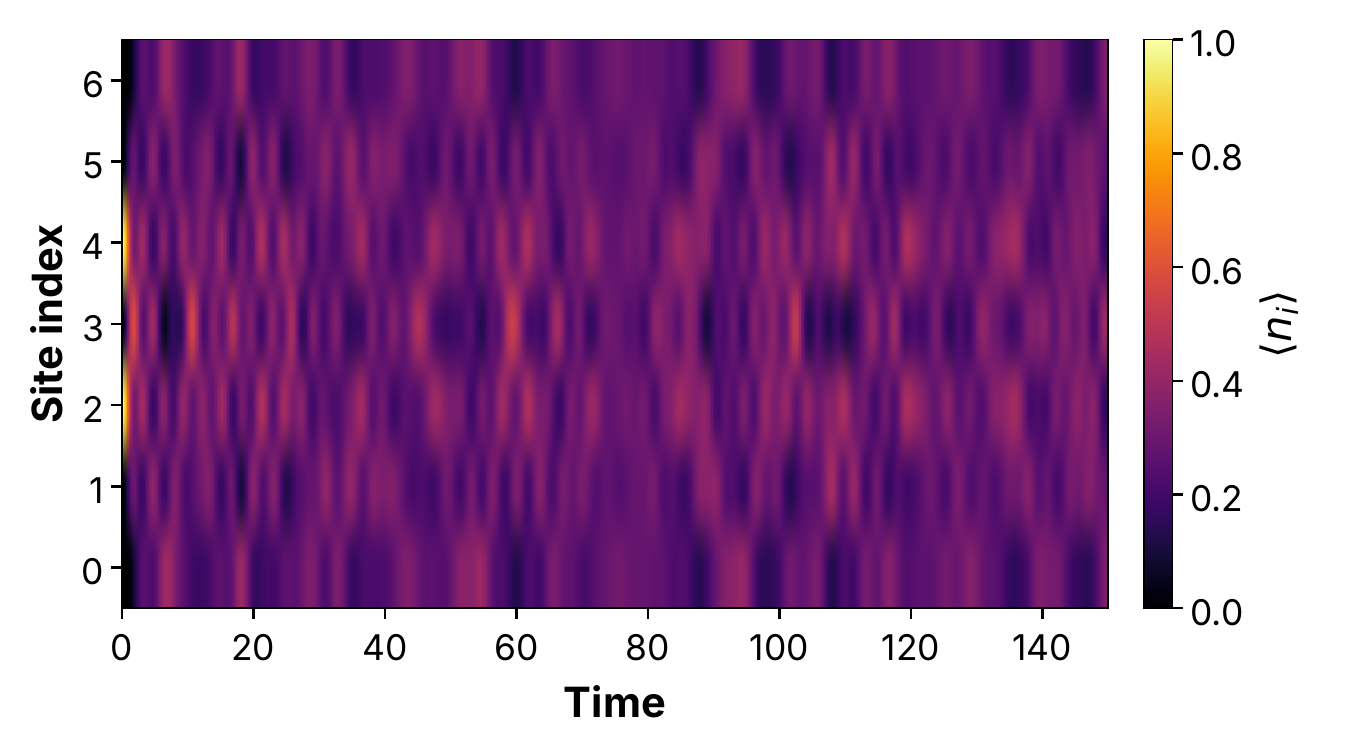}
    \caption{Time evolution of the site occupation $\langle n_i(t)\rangle$ for two initial excitations on an $N=7$ JCH chain with open boundaries, with $g=0.35$, $J=0.5$ and Fock cutoff $d=2$ per qumode. The photons start on sites $2$ and $4$ with all atoms in the ground state, and the evolution is run to $t=150$ with $\Delta t=0.05$ on the matrix-product-state backend. The occupancy stays mirror symmetric about the central site $i=3$ and spreads over the whole chain, with the total $\sum_i\langle n_i\rangle=2$ held to within $10^{-10}$ at all times.}
    \label{fig:jch_multi}
\end{figure}

Figure~\ref{fig:jch_multi} shows the result. At $t=0$ the two excitations are set on sites $2$ and $4$. These spread toward each other and toward the
walls, so the site in between collects weight from both sides. Because
the initial state and the Hamiltonian are both symmetric about site $3$,
the occupancy stays mirror symmetric throughout. Compared to the
single-excitation case the pattern is much less regular, since the
excitations scatter off each other as well as off the boundaries, and at
late times the weight is spread over the chain with most sites close to
$2/7\approx0.29$.

% =====================================================================
\section{Entanglement Entropy dynamics }
\label{sec:entanglement}
% =====================================================================
We calculated the bipartite von Neumann entanglement entropy by factorizing full chain's Hilbert space as $\mathcal{H} = \mathcal{H_A}\otimes\mathcal{H_B}$, where A denotes the local two-qubit block at excited site and B denotes the remaining $N-1$ sites. Because the global dynamics are unitary and initialized from a pure product state, $\ket{\psi(t)}\in \mathcal{H_A}\otimes\mathcal{H_B}$ remains pure for all t. For a bipartition $A\vert B$ of a pure global state, the reduced density matrix $\rho_A(t)=\mathrm{Tr}_B[\ket{\psi(t)}\bra{\psi(t)}]$ is mixed if and only if A and B are quantum mechanically entangled, and by Schmidt decomposition $S(\rho_A)=S(\rho_B)$ identically, independent of the dimensionality of B \cite{Ekert1995}.

At each Trotter step, save\_density\_matrix performs the partial trace over B directly on MPS representation, yielding $\rho_A(t)$ as a $4\times4$ matrix and entropyy is calculated in standard form of,
\begin{equation}
    \operatorname{S}(\rho_A)=-\mathrm{Tr}[\rho_A \log_2\rho_A]=-\Sigma \lambda_i \log_2 \lambda_i
    \label{eq:entropy} 
\end{equation}
where, $\lambda_i$ is the eigenvalues of $\rho_A$, computed via qiskit.quantum\_info.entropy and reporting in base 2 gives the result in bits.

\begin{figure*}[t]
    \centering
    \includegraphics[width=1\linewidth]{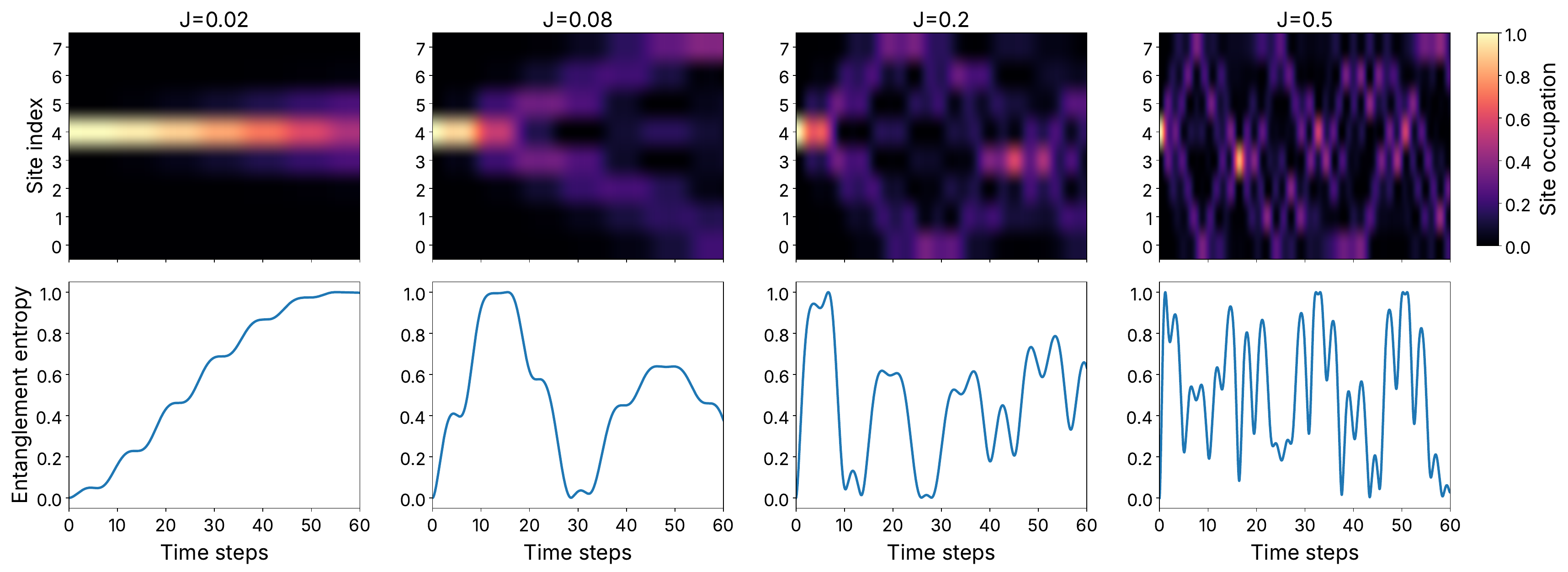}
    \caption{Excitation dynamics of an $N=8$ JCH chain for hopping strengths $J=0.02$, $0.08$, $0.2$ and $0.5$ (left to right), with $g=0.35$ and a single polariton initially at the central site ($i=4$). The top row shows the site occupation $\langle n_i(t)\rangle$ and the bottom row the von Neumann entanglement entropy (in bits) of the central site's reduced two-qubit state with the rest of the chain. Weak hopping gives slow, near-monotonic delocalization, while larger $J$ speeds up the spreading. The results are from Trotterized evolution with $\Delta t=0.05$ up to $t=60$ using Bosonic Qiskit with Fock cutoff $d=2$.}
    \label{fig:Exitation dynamics}
\end{figure*}

The entanglement entropy traces the degree to which the initially localized excitation has become quantum mechanically delocalized across the chain; high entropy (in Figure ~\ref{fig:Exitation dynamics}) signifies that the reduced state at the central site is highly mixed, whereas the low entropy value signifies the excitation remains localized and largely disentangled from the remaining sites. At small value of hopping parameter J=0.02, the induced group velocity is too small for transport to occur within the limited simulated time steps that is why the entropy rises slowly. At J=0.08, the excitation delocalizes much faster than previous and the entropy reaches its maximum at $t\approx 15$. Then,it takes return path to zero again ($t\approx 30$) as the excitation is completely off from the central site. At J=0.2, the dynamics is similar to previous at first but spreading and entropy surges much faster  as excitation delocalizes rapidely. After hitting the chain's boundaries the excitation refocuses partially to the central site. At J=0.5, which now exceeds the JC coupling rate g, the transport resembles a finite size quantum walk; round trips of entropy recur many times and their interference produces the dense,quasi-periodic fringe pattern in the entropy trace.\\


% =====================================================================
\section{Conclusion and outlook}
\label{sec:conclusion}
% =====================================================================

In this work, we demonstrated a computational framework for simulating the real-time dynamics of the one-dimensional Jaynes-Cummings-Hubbard model using Bosonic Qiskit and Matrix Product State (MPS) backends. By tracking a single polariton through an $8$ site cavity chain, our Trotterized simulations successfully captured ballistic wavepacket spreading, boundary reflections, and finite-size revivals across thousands of time steps. Furthermore, by systematically comparing our circuit-based approach against exact diagonalization, we confirmed that discretization and Trotter errors remain tightly bounded. We also extended this setup to the two-excitation sector and found that the framework captures interaction-induced polariton scattering and photon-blockade effects. Finally, we evaluated the bipartite entanglement entropy for different inter-cavity hopping rates. Together, these results show that hybrid qubit-boson software tools can be checked against exact solutions on light-matter lattice models, which is a first step before using them in regimes where no exact reference is available.\\
However, our simulations are restricted to small, one-dimensional open chains in the low-excitation sector with a Fock cutoff of $d = 2$, which enforces a hard-core photon limit and keeps entanglement modest enough for MPS compression. Also, we use only a first-order Trotter decomposition, so the circuit depth accumulates over long evolution times. We leave these for future work. Similarly, several interesting directions remain open. This hybrid simulation framework can be extended to explore richer quantum many-body phenomena and hardware-relevant regimes. A natural next step is to move beyond the hard-core limit into multi-polariton regimes, where photon blockade and effective on-site interactions drive Mott-insulator-to-superfluid phase transitions across larger 2D lattice geometries~\cite{PhysRevB.40.546,Greentree_2006,Hartmann_2006,PhysRevA.76.031805,PhysRevLett.103.086403}. Additionally, incorporating open-system dynamics such as cavity photon loss and atomic spontaneous emission via Kraus operators or Lindblad master equations would allow for benchmarking real-world hardware noise and bosonic quantum error correction codes in superconducting circuit architectures~\cite{Gottesman2001,Albert2018,Cai2021}. Finally, compiling these continuous-variable Trotterized circuits onto real continuous-variable hardware or hybrid oscillator--qubit quantum processing units (QPUs) could unlock scalable digital--analog quantum simulations of complex light--matter systems~\cite{javadiabhari2024quantum,Liu_2026,s5bn-z4jq,selvam2026firstsecondorderdigitalquantum,decker2025symbolichamiltoniancompilerhybrid}.

\acknowledgments
We would like to express our sincere gratitude to Tri-Chandra Multiple Campus, Central Department of Physics (CDP), Tribhuvan University and ICTP PWF Holographic Himalaya for providing a supportive academic environment during the course of this work. We are deeply thankful to Dr. Kiran Adhikari for his continuous encouragement, insightful discussions, and guidance throughout our research pursuits.

\bibliographystyle{apsrev4-2}
\bibliography{ref1.bib}

\clearpage
\onecolumngrid

\appendix

\end{document}